\documentclass[12pt]{article}
\usepackage{graphicx}
\usepackage{amssymb,epsfig}
\usepackage{lscape,graphics,amsmath}
\usepackage{latexsym,amsfonts}
\usepackage{axodraw}
\usepackage[english]{babel}

\newcounter{diags1}
\newcounter{diags2}
\newcounter{diags3}
\begin{document}

\begin{center}
{\Large \bf Neutrino oscillation processes in quantum
field-theoretical approach}\\
\vspace{4mm} Vadim O. Egorov$^{1,2}$, Igor P.~Volobuev$^1$\\
\vspace{4mm} $^1$Skobeltsyn Institute of Nuclear Physics, Moscow State
University
\\ 119991 Moscow, Russia\\
$^2$Faculty of Physics, Moscow State University, 119991 Moscow, Russia
\end{center}

\vspace{0.5cm}
\begin{abstract}
It is shown that neutrino oscillation processes can be
consistently described in the framework of quantum field theory.
Namely, the oscillating electron survival probabilities in
experiments with neutrino detection by charged-current and
neutral-current interactions are calculated in the quantum
field-theoretical approach to neutrino oscillations based on a
modification of the Feynman propagator. The approach is  most
similar to the standard Feynman diagram technique in the momentum
representation. It is found that the oscillating
distance-dependent probabilities of detecting an electron in
experiments with neutrino detection by charged-current and
neutral-current interactions exactly coincide with the
corresponding probabilities calculated in the standard approach.
\end{abstract}

\section{Introduction}
Neutrino oscillation is a well-known and experimentally confirmed
phenomenon, which is usually understood as the transition from a
neutrino flavor state to another neutrino flavor state depending
on the distance traveled
\cite{Giunti:2007ry,Bilenky:2010zza,Petcov}. However, the
situation with the theoretical explanation of this phenomenon is
paradoxical: the phenomenon, which  is both quantum and
relativistic, cannot be consistently described in the framework of
quantum field theory, which is a synthesis of quantum mechanics
and special theory of relativity. The standard theoretical
description of this phenomenon based on the notion of neutrino
flavor states is not perfect: the neutrino flavor states are
superpositions of the neutrino mass eigenstates, and for this
reason the processes with the flavor states cannot be consistently
described within quantum field theory. The problem is the
violation of energy-momentum conservation in such processes,
because in local quantum field theory, where the four-momentum is
conserved in any interaction vertex,  different mass-eigenstate
components of the flavor states must have different momenta as
well as different energies. This problem  was repeatedly discussed
in the literature (see, e.g.
\cite{Giunti:1993se,Grimus:1996av,Naumov:2010um,Lobanov:2015esa}).

A possible solution to the problem with the violation of
energy-momentum conservation is to go off the mass shell. It was
first discussed  in paper \cite{Giunti:1993se}, where it was
suggested that the produced neutrino mass eigenstates are virtual
and their motion to the detection point should be described by the
Feynman propagators.  Later this approach was developed in papers
\cite{Grimus:1996av,Naumov:2010um}. However, the calculations in
these papers imply the use of wave packets and are essentially
different from the standard calculations in the framework of  the
Feynman diagram technique in the momentum representation. This is
due to the standard S-matrix formalism of QFT used in these
papers, which is not appropriate for describing processes at
finite distances and finite time intervals.

In the present paper we will show that neutrino oscillation can be
consistently described in the framework of quantum field theory.
Namely, we will explicitly calculate the probabilities of the
neutrino oscillation processes in experiments with neutrino
detection by charged-current and neutral-current interactions
within a modified perturbative S-matrix formalism, which enables
one   to calculate the amplitudes of the processes passing at
finite distances and finite time intervals. This formalism was put
forward in paper \cite{Volobuev:2017izt}. It is based on the
Feynman diagram technique in the coordinate representation
\cite{Feynman:1949zx} supplemented by new modified rules of
passing to the momentum representation, which will be discussed
below in detail.

\section{Oscillations in experiments with neutrino detection by charged-current interaction}

In the framework of the minimal extension of the Standard Model
(SM) by the right neutrino singlets we consider the case, where
the neutrinos are produced and detected in the charged-current
interaction with  nuclei.  After the diagonalization of the terms
sesquilinear in the neutrino fields, the charged-current
interaction Lagrangian of leptons takes the form
\begin{equation}\label{L_cc}
L_{cc} = - \frac{g }{2\sqrt{2}}\left(\sum_{i,k = 1}^3 \bar l_i
\gamma^\mu (1 - \gamma^5)U_{ik}\nu_k   W^{-}_\mu + h.c.\right),
\end{equation}
where $l_i$ denotes the field of the charged lepton of the i-th
generation, $\nu_i$ denotes the field of the neutrino mass
eigenstate most strongly coupled to $l_i$ and $U_{ik}$ stands for
the Pontecorvo-Maki-Nakagawa-Sakata (PMNS) matrix. Due to this
structure of the interaction Lagrangian any process involving the
production of a neutrino at one point and its detection at another
point, when treated perturbatively, can be represented in the
lowest order by the  following diagram, \vspace*{0.5cm}
\begin{center}
\begin{picture}(193,87)(0,0)
\Text(70.0,94.0)[l]{$e^+ ( q )$}\ArrowLine(67.5,88.0)(40.5,64.5)
\Text(33.5,65.5)[r]{$x$} \Photon(13.5,41.0)(40.5,64.5){2}{3.0}
\Text(53.5,48.5)[r]{$W^+$} \Vertex (13.5,41.0){5} \Vertex
(40.5,64.5){2} \ArrowLine(40.5,64.5)(167.5,64.5) \Vertex
(167.5,64.5){2} \Text(104.8,70.5)[b]{$\nu_i ( p_n )$}
\ArrowLine(167.5,64.5)(194.5,88.0) \Text(197.5,94.0)[l]{$e^- ( k )$}
\Text(175.0,64.5)[l]{$y$} \Photon(167.5,64.5)(194.5,41.0){2}{3.0}
\Vertex (194.5,41.0){5} \Text(170.0,48.5)[r]{$W^+$}
\Text(330.0,60.5)[b]{\addtocounter{equation}{1}(\arabic{equation})}
\setcounter{diags1}{\value{equation}}\label{diag1}
\end{picture}
\end{center}
\vspace*{-1cm} which should be  summed over all three neutrino
mass eigenstates. To be specific, we  assume that the virtual
$W$-bosons are produced and absorbed in interactions with nuclei.
Namely, we suppose that a nucleus  $^{A_1}_{Z_1} X$ that will be
called  nucleus $1$ radiates $W^+$-boson and turns into the
nucleus $^{A_1}_{Z_1 - 1} X$ that will be called nucleus $1^\prime
$, and a nucleus  $^{A_2}_{Z_2} X$ that will be called nucleus $2$
absorbs $W^+$-boson and turns into the nucleus $^{A_2}_{Z_2 + 1}
X$ that will be called nucleus $2^\prime $. Thus, the filled
circles stand for  the matrix elements of the charged weak hadron
current
$$j_\mu^{(1)} = \left <^{A_1}_{Z_1 - 1} X \right|
j_\mu^{(h)} \left| ^{A_1}_{Z_1} X \right>, \quad j_\rho^{(2)} =
\left <^{A_2}_{Z_2 + 1} X \right| j_\rho^{(h)} \left| ^{A_2}_{Z_2}
X \right>,$$ associated with  nuclei $1, 1^\prime$ and $2,
2^\prime$. As it is customary in QFT, we assume that the incoming
nuclei 1 and 2 have definite momenta. Therefore all the three
virtual neutrino eigenstates and the outgoing particles and nuclei
also have definite momenta. In what follows, a 4-momentum of the
virtual neutrino mass eigenstates defined only by the
energy-momentum conservation in the production vertex   will be
denoted by $p_n$ and the one selected also by the experimental
setting will be denoted by $p$.

The amplitude in the coordinate representation corresponding to
diagram (\ref{diag1}) can be  written out in the standard way
using the Feynman rules formulated in  textbook \cite{BOSH}.
According to the prescriptions of the standard perturbative
S-matrix theory (\cite{BOSH}, \S 24), in order to obtain the
amplitude in the momentum representation next we would have to
integrate it with respect to $x$ and $y$ over the Minkowski space.
However, in this case we would get the amplitude of the process
lasting an infinite amount of time and lose the information about
the distance between the production and detection points defined
by the experimental setting. In order to retain this information,
we have to integrate with respect to $x$ and $y$ in such a way
that the distance between these points along the direction of the
neutrino propagation remains fixed. Of course, this is at variance
with the standard S-matrix formalism. However, we recall that the
diagram technique in the coordinate representation was developed
by R.~Feynman \cite{Feynman:1949zx} without reference to S-matrix
theory. Thus, the Feynman diagrams in the coordinate
representation make sense beyond this theory, and for this reason
we can integrate with respect to $x$ and $y$ in any way depending
on the physical problem at hand. In particular, in the case under
consideration we have to integrate in such a way that the distance
between the points $x$ and $y$ along the direction of the
propagation of  neutrino with momentum $\vec p$  defined by the
experimental setting equals to $L$.  This can be achieved by
introducing the delta function $\delta(\vec p(\vec y -\vec
x)/|\vec p| - L)$ into the integral, which is equivalent to
replacing the standard Feynman propagator of the neutrino mass
eigenstate $\nu_i$ in the coordinate representation $S^c_i(y - x)$
by $S^c_i(y - x) \delta(\vec p(\vec y -\vec x)/|\vec p| - L)$. The
Fourier transform of the latter expression was called in paper
\cite{Volobuev:2017izt} the distance-dependent propagator of the
neutrino mass eigenstate $\nu_i$ in the momentum representation.
It will be denoted by $S^c_i(p,L)$ and is defined by the integral:
\begin{equation}\label{prop_L_mom}
S^c_i(p,L) = \int dx\, e^{ipx} S^c_i(x)\,  \delta(\vec p\vec
x/|\vec p| - L).
\end{equation}

This integral can be evaluated exactly by the method of contour
integration \cite{Volobuev:2017izt}, and for $\vec p^{\,2} > m_i^2
- p^2$ the result is given by
\begin{equation}\label{prop_L_mom_a}
S^c_i(p,L) =  i\,\frac{\hat p + \vec \gamma \vec
 p\left(1 - \sqrt{1 + \frac{p^2 - m_i^2}{\vec p^{\,2}}}\,\right) + m_i }
 {2\sqrt{\vec p^{\,2} + p^2 - m_i^2}}\, e^{-i\left(|\vec p| -
\sqrt{\vec p^{\,2} + p^2 - m_i^2}\,\right) L} \,.
\end{equation}
(In paper \cite{Volobuev:2017izt} this distance-dependent
propagator was defined by substituting  the dimensionless delta
function $\delta(\vec p\vec x - |\vec p|L)$ into the integral,
which results in  an extra factor  $|\vec p|$ in the denominator
of $S^c_i(p,L)$. Below we will see that the present definition is
more natural.) We emphasize that this distance-dependent fermion
propagator makes sense only for macroscopic distances $L$.

The results of paper \cite{Grimus:1996av} imply that the virtual
particles propagating at macroscopic distances are almost on the
mass shell. This means that $|p^2 - m_i^2|/ \vec p^{\,2} \ll 1$
and we can expand the square roots to the first order in $(p^2 -
m_i^2)/ \vec p^{\,2}$. It is clear that this term can be dropped
everywhere, except in the exponential, where it is multiplied by a
large macroscopic distance $L$.  In this approximation
distance-dependent propagator (\ref{prop_L_mom_a}) takes the
simple form
\begin{equation}\label{prop_L_mom_b}
S^c_i(p,L) = i\, \frac{\hat p + m_i }{ 2 |\vec
p|}\,e^{-i\frac{m_i^2 - p^2}{2|\vec p|} L}\,.
\end{equation}
 It is worth noting that this distance-dependent fermion propagator taken on
the mass shell has no pole and does not depend on the distance,
which is also true for the exact propagator in formula
(\ref{prop_L_mom_a}).

In fact, we have discussed  this distance-dependent propagator in
order to explain better the  motivations for introducing such an
object, because it exactly corresponds to the experimental
situation in neutrino oscillation processes. However, this
distance-dependent propagator is not convenient for calculating
amplitudes, because there is no inverse  Fourier transformation
for the propagator in formula (\ref{prop_L_mom}). It turns out
that a more convenient and a more fundamental object is the
time-dependent propagator of the neutrino mass eigenstates, which
can be defined as the Fourier transform of $S^c_i (x) \delta(x^0 -
T)$. A similar time-dependent scalar field propagator was
introduced in paper \cite{Volobuev:2017izt}. Using the results of
the calculations of the time-dependent scalar field propagator in
this paper one can easily find that the time-dependent fermion
field propagator in the momentum representation is
\begin{equation}\label{prop_T_mom_}
S^c_i(p_n,T) = i\,\frac{\hat p_n -  \gamma^0 \left(p_n^{\,0}
-\sqrt{(p_n^{\,0})^2 + m_i^2 - p_n^2}\,\right)+
m_i}{2\sqrt{(p_n^{\,0})^2 + m_i^2 - p_n^2}}\, e^{i\left(p_n^{\,0}
-\sqrt{(p_n^{\,0})^2 + m_i^2 - p_n^2}\,\right) T} \,.
\end{equation}
The advantage of the  time-dependent  propagator is that there
exists the inverse Fourier transformation of this propagator,
which allows one to retain the standard  rules of the Feynman
diagram technique just by replacing the Feynman propagator by this
time-dependent propagator. For macroscopic time intervals $T$,
i.e. for the particles close to the mass shell, it looks
explicitly like
\begin{equation}\label{prop_T_mom_c}
S^c_i(p_n,T) = i\, \frac{\hat p_n + m_i } { 2
p_n^{\,0}}\,e^{-i\frac{m_i^2 - p_n^2}{ 2 p_n^{\,0}} T}\,.
\end{equation}
In case all the neutrinos  in a beam have the same momentum $p$
defined by the experimental setting we can express the time $T$ in
terms of the distance $L$ and the neutrino speed $|\vec
p|/p^{\,0}$ as $T = L p^{\,0}/ |\vec p| $, neglect the neutrino
mass that is small compared to $\hat p$  and get a
distance-dependent propagator
\begin{equation}\label{prop_L_mom_c}
S^c_i(p,L) = i\, \frac{\hat p }{ 2 p^{\,0}}\,e^{-i\frac{m_i^2 -
p^2}{ 2 |\vec p |} L}\,,
\end{equation}
which  coincides with the above defined   distance-dependent
propagator of neutrinos (\ref{prop_L_mom_b}) in the approximation
of small neutrino masses. In what follows, we will use propagators
(\ref{prop_T_mom_c}), (\ref{prop_L_mom_c})  for describing
neutrino oscillation processes. We also note that the
time-dependent scalar field propagator  is  adequate for
calculating the probabilities of oscillation processes with
massive scalar mesons, where we cannot neglect their masses.

Now we will calculate the amplitude corresponding  to diagram
(\ref{diag1}) in the case, where the time difference $y^0 - x^0$
is fixed and equal to $T$. Since the momentum transfer in both
production and detection processes is small, we can use the
approximation of Fermi's interaction. Then making use of
time-dependent propagator (\ref{prop_T_mom_c}) and keeping the
neutrino masses only in the exponential we can explicitly write
out the amplitude in the momentum representation corresponding to
diagram (\ref{diag1}) summed over all three neutrino mass
eigenstates:
\begin{equation}\label{amp}
M =  - i\frac{{G_F^{\,2} }}{4{p_n^{\,0}}}\sum\limits_{i = 1}^3
{\left| {U_{1i} } \right|^2 e^{-i\frac{m_i^2 - p_n^2}{ 2
p_n^{\,0}} T} } j_\rho ^{(2)} \bar u \left( k \right)\gamma ^\rho
\left( {1 - \gamma ^5 } \right) \hat p_n\gamma ^\mu \left( {1 -
\gamma ^5 } \right)v \left( q \right)j_\mu ^{(1)} .
\end{equation}
Here $j_\mu ^{(1)}$ and $j_\rho ^{(2)}$ stand for the matrix
elements of the charged weak hadron current associated with nuclei
$1, 1^\prime$ and $2, 2^\prime$; $k$, $p_n$ and $q$ are the
4-momenta of the electron, the intermediate virtual neutrinos and
the positron, respectively, and we do not write out explicitly the
fermion polarization indices.

Averaging with respect to the polarizations of the incoming nuclei
and summing over the polarizations of the outgoing particles and
nuclei one gets the expression for the squared amplitude as
follows:
\begin{equation} \label{7}
\left\langle {\left| M \right|^2 } \right\rangle  =
\frac{{4G_F^{\,4} }}{{\left(  p^{\,0}_n  \right)^2 }}W_{\mu \nu
}^{(1)} A^{\mu \nu \rho \sigma } W_{\rho \sigma }^{(2)} \left[ {1
- 4\sum\limits_{\scriptstyle i,k = 1 \hfill \atop \scriptstyle i <
k \hfill}^3 {\left| {U_{1i} } \right|^2 \left| {U_{1k} } \right|^2
\sin ^2 \left( {\frac{{m_i^2  - m_k^2 }}{{4 {p^{\,0}_n} }}T}
\right)} } \right] ,
\end{equation}
where
$$A^{\mu \nu \rho \sigma } = \frac{1}{64} tr \left ( \hat p_n\gamma ^\mu
\left( {1 - \gamma ^5 } \right) \left( \hat q - m \right) \gamma
^\nu \left( {1 - \gamma ^5 } \right) \hat p_n \gamma ^\sigma
\left( {1 - \gamma ^5 } \right) \left( \hat k + m\right) \gamma
^\rho \left( {1 - \gamma ^5 } \right) \right)
$$
(the factor ${\raise0.5ex\hbox{$\scriptstyle 1$}
\kern-0.1em/\kern-0.15em \lower0.25ex\hbox{$\scriptstyle {64}$}}$
is introduced in order to separate the numerical coefficient from
the Lorentz structure  proper), the tensors $ W_{\mu \nu }^{(1)},
\, W_{\rho \sigma }^{(2)}$ characterizing the interaction of
nuclei $1$ and $2$ with virtual $W$-bosons are defined as
\begin{equation}
 W_{\mu \nu }^{(1)}  = \left\langle {j_\mu ^{(1)} \left( {j_\nu ^{(1)} } \right)^ +  } \right\rangle ,
 \qquad W_{\rho \sigma }^{(2)}  = \left\langle {j_\rho ^{(2)} \left( {j_\sigma ^{(2)} } \right)^ +  }
 \right\rangle.
\end{equation}
Here and below the angle brackets denote  the  averaging with
respect to the polarizations of the incoming particles and the
summation over the polarizations of the outgoing particles, i.e.
in the previous formula they denote the  averaging with respect to
the polarizations of nuclei $1, 2$ and the summation over the
polarizations of nuclei $1^\prime, 2^\prime$.

Since we have dropped the neutrino masses in the time-dependent
propagators, we have actually calculated the amplitude in the
approximation of zero neutrino masses. As we have already noted,
for macroscopic time intervals $T$ the virtual neutrinos are
almost on the mass shell and, therefore, the squared momentum of
the virtual neutrinos $p_n^2$ is also of the order of the neutrino
masses squared and can be neglected. In other words, we may
calculate the squared amplitude in the approximation $p_n^2 = 0$.
Direct calculations show that in this approximation the tensor
$A^{\mu \nu \rho \sigma }$ factorizes:
\begin{equation}
A^{\mu \nu \rho \sigma }
 = \left( -g^{\mu \nu} ( {p_nq})+(  p_n^\mu  q^\nu + q^\mu  p_n^\nu)+ i\varepsilon ^{\mu \nu \alpha \beta } p_{n \alpha}  q_\beta
 \right) \left( -g^{\rho \sigma} ( {p_nk})+(  p_n^\rho  k^\sigma + k^\rho  p_n^{\sigma})- i\varepsilon ^{\rho \sigma \alpha \beta } p_{n\alpha}  k_\beta
 \right).
\end{equation}
Correspondingly, the squared amplitude in formula (\ref{7})
factorizes as follows:
\begin{eqnarray} \label{factors}
\left\langle {\left| M \right|^2 } \right\rangle  &=& \left\langle
{\left| M_1 \right|^2 } \right\rangle \left\langle {\left| M_2
\right|^2 } \right\rangle \frac{{1}}{{4 \left( { p^{\,0}_n}
\right)^2 }} \left[ {1 - 4\sum\limits_{\scriptstyle i,k = 1 \hfill
\atop \scriptstyle i < k \hfill}^3 {\left| {U_{1i} } \right|^2
\left| {U_{1k} } \right|^2 \sin ^2 \left( {\frac{{m_i^2  - m_k^2
}}{{4 p^{\,0}_n}}T} \right)} } \right] ,\\
\label{factor1} \left\langle {\left| M_1 \right|^2 } \right\rangle
&=& 4G_F^{\,2} \left( -{g^{\mu \nu } \left( {p_nq} \right) +\left(
{p_n^\mu  q^\nu   + q^\mu  p_n^\nu  } \right) +
i\varepsilon ^{\mu \nu \alpha \beta } p_{n \alpha}  q_\beta  } \right) W_{\mu \nu }^{(1)} , \\
\label{factor2} \left\langle {\left| M_2 \right|^2 } \right\rangle
&=& 4G_F^{\,2} \left(- {g^{\rho \sigma } \left( {p_n k} \right) +
\left( {p_n^\rho  k^\sigma   + k^\rho  p_n^\sigma  } \right) -
i\varepsilon ^{\rho \sigma \alpha \beta } p_{n \alpha}  k_\beta  }
\right) W_{\rho \sigma }^{(2)} .
\end{eqnarray}
Here $\left\langle {\left| M_1 \right|^2 } \right\rangle$ is the
squared amplitude of the decay process of nucleus $1$ into nucleus
$1^\prime$,  positron and a massless fermion and $\left\langle
{\left| M_2 \right|^2 } \right\rangle$ is the squared amplitude of
the process of electron production in the collision of the
massless fermion and nucleus $2$.

Now we are in a position to calculate the probability of the
process depicted in diagram (\ref{diag1}), when the time
difference between the points $x$ and $y$  is equal to $T$. We
will do these calculations in accordance with the rules of the
standard perturbative S-matrix theory, although we are aware that
the rules of calculating the probabilities of processes passing at
finite time interval and finite distances  may be different from
those of the standard S-matrix theory. We will discuss this
difference below. To this end we denote the 4-momenta of the
nuclei by $P^{(i)} = (E^{(i)}, \vec P^{(i)}), \ P^{(i^\prime)} =
(E^{(i^\prime)}, \vec P^{(i^\prime)}), \ i= 1,2$, and recall that
the amplitude in the momentum representation corresponding to
diagram (\ref{diag1}) contains, along with the expression in
formula (\ref{amp}), the delta function of energy-momentum
conservation. Thus, to calculate the probability of the process
per unit time per unit volume, we have to multiply amplitude
(\ref{factors}) by $(2\pi)^4 \delta ( P^{(1)} + P^{(2)} -
P^{(1^\prime)} - P^{(2^\prime)}- q - k)$ and to integrate it with
respect to the momenta of the outgoing particles and nuclei.

Since the momentum $p_n$ of the virtual neutrinos is defined by
the energy-momentum conservation in the production vertex, $p_n =
P^{(1)} - P^{(1^\prime)} - q$, this integration can lead to
variation in the virtual neutrino momentum, which contradicts the
experimental situation in neutrino oscillations, where the virtual
neutrinos propagate in the direction defined by the relative
position of a source and a detector. This means that we have to
calculate the differential probability of the process with $p_n$
fixed by the experimental setting.

Let us denote by $\vec p$ the momentum that is directed from the
source to the detector and satisfies the momentum conservation
condition $\vec P^{(1)} - \vec P^{(1^\prime)} - \vec q - \vec p =
0$ in the production vertex and define the four-momentum  $p = (
p^{\,0}, \vec p), \, p^{\,0} > 0,  \, p^2 =0$. The required
differential probability of the process with $p_n$ fixed can be
obtained by multiplying amplitude  (\ref{factors})  by the delta
function $\delta (p_n- p)$ or, equivalently, by replacing $p_n$ by
$p$ in the amplitude and multiplying it by $\delta ( P^{(1)} -
P^{(1^\prime)} - q - p )$. This is consistent, because we work in
the approximation of massless neutrinos.

Thus, the differential probability takes the form:
\begin{eqnarray}
\frac{dW}{dp}  &=& \frac{{G_F^4 }}{{16\left( {2\pi } \right)^8
E^{(1)} E^{(2)} \left( {p^{\,0}} \right)^2 }}\left[ {1 -
4\sum\limits_{\scriptstyle i,k = 1 \hfill \atop \scriptstyle i < k
\hfill}^3 {\left| {U_{1i} } \right|^2 \left| {U_{1k} } \right|^2
\sin ^2 \left( {\frac{{m_i^2  - m_k^2 }}{{4{p^{\,0}}}}T} \right)} } \right] \times \nonumber \\
& & \times \int {d^3 qd^3 P^{(1')} d^3 kd^3 P^{(2')} \frac{1}{{q^0 E^{(1')} k^0 E^{(2')} }}}  \times  \nonumber \\
& & \times \left( {g^{\mu \nu } \left( {pq} \right) - \left(
{p^\mu  q^\nu   + q^\mu  p^\nu  } \right) - i\varepsilon ^{\mu \nu
\alpha \beta } p_\alpha  q_\beta  } \right)W_{\mu \nu }^{(1)}
\left( {P^{(1)} ,P^{(1')} } \right) \times  \nonumber \\
& & \times \left( {g^{\rho \sigma } \left( {pk} \right) - \left(
{p^\rho  k^\sigma   + k^\rho  p^\sigma  } \right) + i\varepsilon
^{\rho \sigma \alpha \beta } p_\alpha  k_\beta  } \right)W_{\rho
\sigma }^{(2)}
\left( {P^{(2)} ,P^{(2')} } \right) \times  \nonumber \\
& & \times \delta \left( {P^{(1)}  + P^{(2)}  - P^{(1')}  -
P^{(2')}  - q - k} \right)\delta \left( {P^{(1)}  - P^{(1')}  - q
- p} \right) .
\end{eqnarray}
It is easy to verify  that, due to the factorization of the
squared amplitude,  this differential probability also factorizes.
Now, since the momentum of virtual neutrinos is fixed, we can
replace $T$ by $L p^{\,0}/|\vec p|$, as it was explained after
formula (\ref{prop_T_mom_c}), which gives
\begin{equation} \label{fact2}
\frac{{dW}}{{dp}} = \frac{1}{{2\pi }}\frac{{dW_1 }}{{d\vec p}}W_2
\left[ {1 - 4\sum\limits_{\scriptstyle i,k = 1 \hfill \atop
\scriptstyle i < k \hfill}^3 {\left| {U_{1i} } \right|^2 \left|
{U_{1k} } \right|^2 \sin ^2 \left( {\frac{{m_i^2  - m_k^2
}}{{4\left| {\vec p} \right|}}L} \right)} } \right] ,
\end{equation}
where
\begin{eqnarray}
\frac{{dW_1 }}{{d\vec p}} &=& \frac{1}{{2E^{(1)} }}
\frac{1}{{\left( {2\pi } \right)^3 }} \frac{1}{{2p^{\,0} }} \int
{\frac{{d^3 q}} {{\left( {2\pi } \right)^3 }}\frac{1}{{2q^0
}}\frac{{d^3 P^{(1')} }} {{\left( {2\pi } \right)^3
}}\frac{1}{{2E^{(1')} }} \left\langle {\left| {M_1 } \right|^2 }
\right\rangle \left( {2\pi } \right)^4
\delta \left( {P^{(1)}  - P^{(1')}  - q - p} \right)}  =  \nonumber \\
& & \nonumber  \\
&=& \frac{{G_F^2 }}{{4\left( {2\pi } \right)^5 E^{(1)} p^{\,0}
}}\int {d^3 qd^3 P^{(1')} \frac{1}{{q^0 E^{(1')} }}\left( { -
g^{\mu \nu } \left( {pq} \right) + \left( {p^\mu  q^\nu   + q^\mu
p^\nu  } \right) +
i\varepsilon ^{\mu \nu \alpha \beta } p_\alpha  q_\beta  } \right) \times } \nonumber  \\
& & \times W_{\mu \nu }^{(1)} \left( {P^{(1)} ,P^{(1')} }
\right)\delta \left( {P^{(1)}  - P^{(1')}  - q - p} \right)
\end{eqnarray}
is the differential probability of the decay of  nucleus $1$ into
nucleus $1'$, positron and a massless fermion  with momentum $\vec
p$, which coincides with the sum of the differential probabilities
of the decay of  nucleus $1$ into nucleus $1'$, positron and all
the three neutrino mass eigenstates, and
\begin{eqnarray}
W_2  &=& \frac{1}{{2E^{(2)} 2p^{\,0} }}\int {\frac{{d^3
k}}{{\left( {2\pi } \right)^3 }} \frac{1}{{2k^0 }}\frac{{d^3
P^{(2')} }}{{\left( {2\pi } \right)^3 }}\frac{1}{{2E^{(2')} }}
\left\langle {\left| {M_2 } \right|^2 } \right\rangle \left( {2\pi
} \right)^4
\delta \left( {P^{(2)}  + p - P^{(2')}  - k} \right)}  =  \nonumber \\
& & \nonumber \\
&=& \frac{{G_F^2 }}{{4\left( {2\pi } \right)^2 E^{(2)} p^{\,0}
}}\int {d^3 kd^3 P^{(2')} \frac{1}{{k^0 E^{(2')} }}\left( { -
g^{\rho \sigma } \left( {pk} \right) + \left( {p^\rho  k^\sigma
+ k^\rho p^\sigma  } \right) -
i\varepsilon ^{\rho \sigma \alpha \beta } p_\alpha  k_\beta  } \right) \times }  \nonumber \\
& & \times W_{\rho \sigma }^{(2)} \left( {P^{(2)} ,P^{(2')} }
\right)\delta \left( {P^{(2)}  + p - P^{(2')}  - k} \right)
\end{eqnarray}
is the probability of the scattering process  of a massless
fermion with  momentum $\vec p$ and  nucleus $2$ resulting in the
production  of  nucleus $2'$ and an electron, which coincides with
the sum of the probabilities of the scattering processes  of all
the three  neutrino mass eigenstates and  nucleus $2$. The terms
in the square brackets in formula (\ref{fact2}) reproduce the
standard expression for the oscillating neutrino or electron
survival probability.

The physical considerations suggest that the differential
probability  $ \frac{{dW}}{{dp}} $ for $L \to 0$ should be equal
to the product $\frac{{dW_1 }}{{d\vec p}}W_2 $, i.e. there is an
extra factor ${2\pi }$ in the denominator of formula
(\ref{fact2}). This means that the standard rules of calculating
the process probabilities in  perturbative S-matrix theory should
be  modified so as to include the extra factor $2\pi$.

\begin{figure}[h]
\centering
\includegraphics[width=1\linewidth]{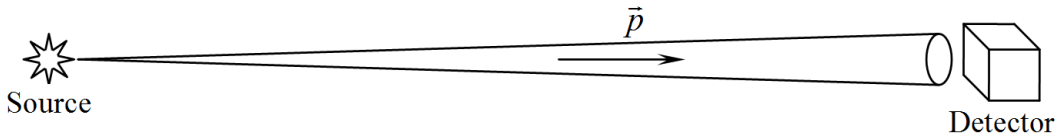}
\caption{Illustration of variance in the direction of the virtual
neutrino momenta due to a non-zero detector size.}
\label{fig-1}
\end{figure}

The appearance of this extra factor can be explained  as follows:
in fact, the detector registers not only the neutrinos with
momentum $\vec p$ from a point-like source, but also the neutrinos
with the momenta, which lie inside a small cone (see Figure
\ref{fig-1}) with the axis along the vector $\vec p$. This is due
to a non-zero size of the detector.
Obviously, the picture has an  approximate circular symmetry about
the direction of  momentum $\vec p$, which gives the factor $2
\pi$ after the integration with respect to  the azimuthal angle.
Thus, the rules of the  standard  perturbative S-matrix theory
should be modified in our case so as to include the factor $2 \pi$
along with the delta function $\delta ( P^{(1)} - P^{(1^\prime)} -
q - p )$, which fixes the 4-momentum $p$ of the intermediate
neutrinos. Correspondingly, the final formula for the differential
probability of the process under consideration  looks like
\begin{eqnarray}
\frac{{dW}}{{dp}} &=& \frac{1}{{2E^{(1)} 2E^{(2)} }} \int {\frac{{d^3 k}}{{\left( {2\pi } \right)^3 }}
\frac{1}{{2k^0 }}\frac{{d^3 q}}{{\left( {2\pi } \right)^3 }}\frac{1}{{2q^0 }}
\frac{{d^3 P^{(1')} }}{{\left( {2\pi } \right)^3 }}\frac{1}{{2E^{(1')} }}
\frac{{d^3 P^{(2')} }}{{\left( {2\pi } \right)^3 }}\frac{1}{{2E^{(2')} }}
\left\langle {\left| M \right|^2 } \right\rangle  \times } \nonumber \\
& & \times \left( {2\pi } \right)^4 \delta \left( {P^{(1)}  +
P^{(2)}  - P^{(1')}  - P^{(2')}  - q - k} \right) 2\pi \delta
\left( {P^{(1)}  - P^{(1')}  - q - p} \right),
\end{eqnarray}
which eliminates the contradiction and provides the consistent
result: $ \left. {\frac{{dW}}{{dp}}} \right|_{L = 0}  =
\frac{{dW_1 }}{{d\vec p}}W_2  $.

In the approximation of massless neutrinos  $\frac{{dW_1 }}{{d\vec
p}} $ coincides with the neutrino probability flux and $W_2  $
coincides with the cross section of the scattering process of a
massless fermion on nucleus 2. Thus, we have obtained that the
probability of detecting an electron is equal to the probability
of the neutrino production in the source multiplied by the
probability of the neutrino interaction in the detector and the
standard distance-dependent electron or neutrino survival
probability, i.e. we have actually exactly reproduced the result
of the standard approach to neutrino oscillations in the framework
of QFT without making use of the neutrino flavor states. It is
necessary to note that this result differs from the results of
papers
\cite{Giunti:1993se,Grimus:1996av,Naumov:2010um,Lobanov:2015esa}
using the standard perturbative S-matrix theory, because all these
papers reproduce the result of the standard approach with various
corrections. If nuclei 1 in the source have a momentum
distribution, the total neutrino probability flux can be obtained
by performing the average of $\frac{{dW_1 }}{{d\vec p}} $ over the
momentum distribution of nucleus 1, and the number of events in
the detector per unit time can be found by integrating the
corresponding  differential probability $ \frac{{dW }}{{dp}} $ and
the densities of nuclei 1 and nuclei 2 over the volumes of the
neutrino source and detector.

The rules  for calculating the probabilities of neutrino
oscillation processes passing at finite distances and finite time
intervals were suggested by the factorized structure of their
squared amplitudes arising due to the extremely small neutrino
masses. However, we believe that  these rules can be used for
calculating the probabilities of any processes passing at finite
distances and finite time intervals. In particular, they can  also
be used for calculating the probabilities of oscillation processes
with neutral kaons, where the differential probability of the
processes factorizes exactly like in formula (\ref{fact2}) due to
a simpler structure of the amplitude of such processes in the case
of scalar particles.

\section{Oscillations in experiments with neutrino detection by neutral-current and charged-current interactions}
Now we consider the case, where the neutrinos are produced  in the
charged-current interaction with  nuclei and detected in both
neutral-current and  charged-current interactions with electrons,
as it is done in the Kamiokande experiment. The corresponding
processes are described by the following Feynman diagrams:
\vspace{2cm}
\begin{center}
\begin{picture}(193,81)(0,0)
\Text(70.0,94.0)[l]{$e^+ ( q)$}\ArrowLine(67.5,88.0)(40.5,64.5)
\Text(33.5,65.5)[r]{$x$} \Photon(13.5,41.0)(40.5,64.5){2}{3.0}
\Text(53.5,48.5)[r]{$W^+$} \Vertex (13.5,41.0){5} \Vertex
(40.5,64.5){2} \ArrowLine(40.5,64.5)(167.5,64.5) \Vertex
(167.5,64.5){2} \Text(104.8,70.5)[b]{$\nu_i ( p_n )$}
\ArrowLine(167.5,64.5)(194.5,88.0) \Text(198.0,94.0)[l]{$\nu_i ( k_2 )$}
\Text(175.0,64.5)[l]{$y$} \Text(166.0,48.5)[l]{$Z$}
\Photon(167.5,64.5)(194.5,41.0){2}{3.0} \Vertex (194.5,41.0){2}
\ArrowLine(167.5,17.5)(194.5,41.0) \Text(132.5,17.5)[l]{$e^- ( k_1 )$}
\ArrowLine(194.5,41.0)(221.5,17.5) \Text(225.0,17.5)[l]{$e^- ( k )$}
\Text(330.0,60.5)[b]{\addtocounter{equation}{1}(\arabic{equation})}
\setcounter{diags2}{\value{equation}}
\end{picture}
\end{center}

\vspace{1cm}
\begin{center}
\begin{picture}(193,81)(0,0)
\Text(70.0,94.0)[l]{$e^+ (q)$}\ArrowLine(67.5,88.0)(40.5,64.5)
\Text(33.5,65.5)[r]{$x$} \Photon(13.5,41.0)(40.5,64.5){2}{3.0}
\Text(53.5,48.5)[r]{$W^+$} \Vertex (13.5,41.0){5} \Vertex
(40.5,64.5){2} \ArrowLine(40.5,64.5)(167.5,64.5) \Vertex
(167.5,64.5){2} \Text(104.8,70.5)[b]{$\nu_k ( p_n )$}
\ArrowLine(167.5,64.5)(194.5,88.0) \Text(197.5,94.0)[l]{$e^- ( k )$}
\Text(175.0,64.5)[l]{$y$} \Text(160.0,48.5)[l]{$W^+$}
\Photon(167.5,64.5)(194.5,41.0){2}{3.0} \Vertex (194.5,41.0){2}
\ArrowLine(167.5,17.5)(194.5,41.0) \Text(132.5,17.5)[l]{$e^- ( k_1 )$}
\ArrowLine(194.5,41.0)(221.5,17.5) \Text(225.0,17.5)[l]{$\nu_i ( k_2 )$}
\Text(330.0,60.5)[b]{\addtocounter{equation}{1}(\arabic{equation})}
\setcounter{diags3}{\value{equation}}
\end{picture}
\end{center}
It is clear that in calculating the amplitude of the process the
contribution of  diagram (\arabic{diags3}) should be taken with
all three neutrino mass eigenstates, i.e. $k = 1,2,3.$ Since only
the final electron is registered experimentally, the probabilities
of the processes with different final neutrino states should be
summed up to give the  probability of registering an electron.

Now let us denote the particle momenta as follows: the momentum of
the positron is $q$, the momentum of the virtual neutrinos is
$p_n$, the momentum of the outgoing electron is $k$, the momentum
of the incoming electron  is $k_1$ and the momentum of the
outgoing neutrino is $k_2$. Again we use the approximation of
Fermi's interaction and take time-dependent propagator
(\ref{prop_T_mom_c}) keeping the neutrino masses only in the
exponential. Then the amplitude corresponding to diagram
(\arabic{diags2}) in the momentum representation looks like
\begin{eqnarray}
\label{amp_nc} M_{nc}^{(i)} &=& i \frac{{G_F^{\,2} }}{4p_n^0}\, { U_{1i} ^* } e^{
- i\frac{{m_i^2  - p_n^2 }}{{2 { p^{\,0}_n} }}T}\, \bar \nu_i
\left( k_2 \right)\gamma ^\mu \left( {1 - \gamma ^5 } \right) \hat
p_n\gamma ^\rho \left(
{1 - \gamma ^5 } \right) v \left( q \right)j_\rho \times \\
 & & \times \left[ \left( -\frac{1}{2} + \sin^2 \theta_W \right)
\bar u \left( k \right)\gamma _\mu (1 - \gamma^5) u \left( k_1
\right) + \sin^2 \theta_W \bar u \left( k \right)\gamma _\mu (1 +
\gamma^5) u \left( k_1 \right)\right]. \nonumber
\end{eqnarray}
Similarly, the sum over $k $ of the amplitudes corresponding to
diagram (\arabic{diags3}) can be written out to be
\begin{eqnarray}
M_{cc}^{(i)} & = &-i \frac{{G_F^{\,2} }}{4{p_n^0}}\, {U_{1i}^*
}\left( \sum\limits_{k = 1}^3 {\left| {U_{1k} } \right|^2} e^{ -
i\frac{{m_i^2  - p_n^2 }}{{2 { p^{\,0}_n} }}T} \right) \bar u
\left( k \right)\gamma ^\mu (1 - \gamma^5) \hat p_n\gamma ^\rho
\left( {1 - \gamma ^5 } \right)
v\left( q \right) j_\rho \times \nonumber \\
\label{amp_cc} & & \times \bar \nu_i \left(
k_2 \right)\gamma _\mu (1 - \gamma^5) u \left( k_1 \right).
\end{eqnarray}
Next it is convenient to use  the Fierz identity, which transposes
the spinors $\bar u \left( k \right)$ and  $\bar \nu_i \left( k_2
\right)$ in the amplitude $M_{cc}^{(i)}$ and makes this amplitude
look similar to $M_{nc}^{(i)}$, and  to introduce the following
notations for the time-dependent factors:
\begin{equation}\label{factors_dd}
A_i = { U_{1i}^* } e^{ - i\frac{{m_i^2  - p_n^2 }}{{2 { p^{\,0}_n}
}}T}, \quad B_i = {U_{1i}^* }\left( \sum\limits_{k = 1}^3 {\left|
{U_{1k} } \right|^2} e^{ - i\frac{{m_i^2  - p_n^2 }}{{2 {
p^{\,0}_n} }}T} \right).
\end{equation}
Then the total amplitude of the process with neutrino $\nu_i$ in
the final state, which is  the sum of the amplitudes
$M_{nc}^{(i)}$ and $M_{cc}^{(i)}$, can be represented as follows:
\begin{eqnarray}\label{amp_tot}
 M_{tot}^{(i)} = i \frac{{G_F^{\,2} }}{4p_n^{\,0}}\,  \bar \nu_i \left( k_2 \right)\gamma ^\mu \left( {1 -
\gamma ^5 } \right) \hat p_n\gamma ^\rho \left( {1 - \gamma ^5 }
\right) v \left( q \right)j_\rho \times  \phantom{aaaaaaaaaaaaaaaaaaaaaaaaaaaaaa} \\
\nonumber  \times \left[ \left(B_i + A_i \left(-\frac{1}{2} + \sin^2
\theta_W \right)\right) \bar u \left( k \right)\gamma _\mu (1 -
\gamma^5) u \left( k_1 \right) + A_i \sin^2 \theta_W \bar u \left(
k \right)\gamma _\mu (1 + \gamma^5) u \left( k_1 \right)\right].
\end{eqnarray}
Now we have to calculate the squared amplitude, averaged with
respect to the polarizations of the incoming nucleus and particles
and summed over the polarizations of the outgoing nucleus and
particles. Similar to the case of the neutrino detection in
charged-current interaction, in the approximation $p_n^2 = 0$ the
squared amplitude  factorizes as follows:
\begin{eqnarray} \label{factors_nc}
\left\langle {\left| M_{tot}^{(i)} \right|^2 } \right\rangle  &=&
\left\langle {\left| M_1 \right|^2 } \right\rangle \left\langle
{\left| M_2^{(i)} \right|^2 } \right\rangle \frac{{1}}{{4 (p_n^{\,0})^2 }},\\
\label{factor1_nc} \left\langle {\left| M_1 \right|^2 }
\right\rangle &=& 4 G_F^{\,2} \left( -{g^{\mu \nu } \left( {p_nq}
\right) + \left( {p_n^\mu  q^\nu  +  q^\mu  p_n^\nu  } \right) +
i\varepsilon ^{\mu \nu \alpha \beta } p_{n\alpha}  q_\beta  } \right)W_{\mu \nu }^{(1)}, \\
\left\langle {\left| M_2^{(i)} \right|^2 } \right\rangle &=& 64
G_F^{\,2}\left[\left|B_i + A_i \left(-\frac{1}{2} + \sin^2
\theta_W \right)\right|^2 (p_nk_1)^2 +
\left|A_i\right|^2 \sin^4 \theta_W  (p_nk)^2 - \right. \nonumber \\
\label{factor2_nc} & & - \sin^2 \theta_W Re \left( \left(B_i + A_i
\left(-\frac{1}{2} + \sin^2 \theta_W \right)\right) A_i^* \right)
(p_nk_2)m^2 \Bigg].
\end{eqnarray}
Here $\left\langle {\left| M_1 \right|^2 } \right\rangle$ is the
squared amplitude of the decay process of nucleus $1$ into nucleus
$1^\prime$,  positron and a massless fermion, $W_{\mu \nu
}^{(1)}$  denoting the corresponding averaged product of the
matrix elements of the charged weak hadron current, and
$\left\langle {\left| M_2^{(i)} \right|^2 } \right\rangle$ is the
squared amplitude of the scattering process of  a massless fermion
and the incoming electron.

As we have found in the previous section for the case of neutrino
registration in charged-current interaction, to obtain the
differential probability of the process we have to multiply the
amplitude $\left\langle {\left| M_{tot}^{(i)} \right|^2 }
\right\rangle$ by the delta function of energy-momentum
conservation $(2\pi)^4\delta ( P^{(1)} + k_1 - P^{(1^\prime)} - q
- k - k_2)$ and by the delta function  $2\pi \delta ( P^{(1)} -
P^{(1^\prime)} - q - p )$ that selects the momentum of the virtual
neutrinos, substitute $p$ instead of $p_n$ in it and to integrate
with respect to the momenta of the outgoing particles and nucleus.
This gives
\begin{eqnarray} \label{prob_nc}
\frac{d W^{(i)}}{d p} &=& \frac{d W_1}{d \vec p}\, W_2^{(i)}, \\
\label{dec_prob_nc} \frac{d W_1}{d \vec p} &=& \int
\frac{\left\langle {\left| M_1 \right|^2 }
\right\rangle}{{2E^{(1)} }} (2 \pi)^4 \delta  \left( {P^{(1)}  -
P^{(1')}  - q - p} \right)\frac{{d^3 q}}{{\left( {2\pi } \right)^3
}{2q^0 }} \frac{{d^3 P^{(1')} }}{{\left( {2\pi } \right)^3
}{2E^{(1')}}} \frac{1}{{\left( {2\pi } \right)^3 }{2 p^{\,0} }}, \\
\label{scat_prob_nc} W_2^{(i)}  &=& \int \frac{\left\langle
{\left| M_2^{(i)} \right|^2 } \right\rangle}{2 p^{\,0} 2k_1^0
}(2\pi)^4\delta ( k_1 + p - k - k_2) \frac{{d^3 k}}{{\left( {2\pi
} \right)^3 }{2k^0 }} \frac{{d^3 k_2}}{{\left( {2\pi } \right)^3
}{2k_2^0 }}.
\end{eqnarray}
Formula (\ref{dec_prob_nc}) means that $\frac{d W_1}{d \vec p}$ is
the differential decay probability of nucleus 1 into nucleus
$1^\prime$, positron and a massless neutral fermion with fixed
momentum $\vec p$. In fact, this is the sum of the differential
decay probabilities of nucleus 1 into nucleus $1^\prime$, positron
and all the three neutrino mass eigenstates $\nu_i$ with fixed
momentum $\vec p$ taken in the approximation of zero neutrino
masses. Similarly, in accordance with formula
(\ref{scat_prob_nc}), the probability $W_2^{(i)}$ is the
probability of the process of scattering of a massless neutral
fermion and electron  with the production of electron and the
neutrino mass eigenstate $\nu_i$ in the final state. Obviously,
this probability is the sum of the probabilities of the processes
of scattering of
 electron and all the three neutrino mass eigenstates
 with the production of electron and the neutrino mass eigenstate $\nu_i$ in the
final state.

To obtain the probability of finding an electron in the final
state we have to sum the probability $\frac{d W^{(i)}}{d p}$ over
$i = 1,2,3$. Obviously, this reduces to summing over $i$ the
squared amplitude $ \left\langle {\left| M_2^{(i)} \right|^2 }
\right\rangle$, because only this amplitude  depends on $i$.

Since now the virtual neutrinos have  fixed momentum $p$, we can
replace $T$ by $L p^{\,0}/|\vec p|$ in all the subsequent
formulas. Then the definition of the time-dependent factors $A_i$
and $B_i$ in (\ref{factors_dd}) leads to the following expressions
for their absolute values and  products:
\begin{eqnarray} \label{products_dd}
\left|A_i \right|^2 &=& \left|U_{1i} \right|^2, \\
\left|B_i \right|^2 &=& \left|U_{1i} \right|^2 \left[ {1 -
4\sum\limits_{\scriptstyle k,l = 1 \hfill \atop \scriptstyle  k <
l \hfill}^3 {\left| {U_{1k} } \right|^2 \left| {U_{1l} } \right|^2
\sin ^2 \left( {\frac{{m_k^2  - m_l^2 }}{{4\left| {\vec p}
\right|}}L} \right)} } \right], \\
Re \left(A_i  B_i^* \right) &=& \left|U_{1i} \right|^2
\sum\limits_{\scriptstyle k = 1 }^3 \left| {U_{1k} } \right|^2
\cos ^2 \left( {\frac{{m_k^2  - m_i^2 }}{{4\left| {\vec p}
\right|}}L} \right).
\end{eqnarray}
Substituting these expressions into formula (\ref{factor2_nc}) and
summing over $i$, we get
\begin{eqnarray}
 & & \sum\limits_{\scriptstyle i = 1 }^3 \left\langle {\left|
M_2^{(i)} \right|^2 }
\right\rangle  = 64 G_F^{\,2}\Bigg[ \sin^4 \theta_W  (pk)^2 +  \\
& & + \left(\left(\frac{1}{2} + \sin^2 \theta_W \right)^2 - 8 \sin^2 \theta_W
\sum\limits_{\scriptstyle k,l = 1 \hfill \atop \scriptstyle  k < l
\hfill}^3 {\left| {U_{1k} } \right|^2 \left| {U_{1l} } \right|^2
\sin ^2 \left( {\frac{{m_k^2 - m_l^2 }}{{4\left| {\vec p}
\right|}}L} \right)}  \right)(pk_1)^2
- \nonumber \\
& & \left.  - \left( \sin^2
\theta_W\left(\frac{1}{2} + \sin^2 \theta_W \right) - 4 \sin^2
\theta_W \sum\limits_{\scriptstyle k,l = 1 \hfill \atop
\scriptstyle  k < l \hfill}^3 \left| {U_{1k} } \right|^2 \left|
{U_{1l} } \right|^2 \sin ^2 \left( {\frac{{m_k^2 - m_l^2
}}{{4\left| {\vec p} \right|}}L} \right) \right) (pk_2)m^2
\right]. \nonumber
\end{eqnarray}
Next substituting this expression into formula
(\ref{scat_prob_nc}) summed over $i$ and evaluating the integrals
with the help of the formulas for neutrino-electron scattering
kinematics presented in \S 16 of textbook \cite{Okun:1982ap}, we
arrive at  the following result:
\begin{eqnarray}
W_2  &=& \frac{{G_F^2 m}}{{2\pi }}\frac{{2\left| {\vec p} \right|^2 }}{{2\left| {\vec p} \right| + m}}
\left[ {1 - 2\sin ^2 \theta _W \left( {1 + \frac{{2\left| {\vec p} \right|}}{{2\left| {\vec p} \right| + m}}} \right) +
4\sin ^4 \theta _W \left( {1 + \frac{1}{3}\left( {\frac{{2\left| {\vec p} \right|}}{{2\left| {\vec p} \right| + m}}} \right)^2 } \right) + } \right. \nonumber \\
& & \left. { + 4\sin ^2 \theta _W \left( {1 + \frac{{2\left| {\vec
p} \right|}}{{2\left| {\vec p} \right| + m}}} \right)\left( {1 -
4\sum\limits_{\scriptstyle k,l = 1 \hfill \atop \scriptstyle k < l
\hfill}^3 {\left| {U_{1k} } \right|^2 \left| {U_{1l} } \right|^2
\sin ^2 \left( {\frac{{m_k^2  - m_l^2 }}{{4\left| {\vec p}
\right|}}L} \right)} } \right)} \right].
\end{eqnarray}
We see that in the case, where the neutrinos are produced  in the
charged-current interaction with  nuclei and detected in both
neutral-current and charged-current interactions with electrons,
the situation is different from the case, where neutrinos are
detected in charged-current interactions. The same situation takes
place in the standard approach and  the formula for $W_2$ exactly
coincides with the expression $P_{\nu_e \to \nu_e} \sigma_{\nu_e
e} + ( 1 - P_{\nu_e \to \nu_e}) \sigma_{\nu_\mu e} $, which one
expects for this quantity in the standard approach.

\section{Conclusion}
In the present paper we have shown that it is possible to
calculate consistently neutrino oscillation processes in a quantum
field-theoretical approach within the framework of the SM
minimally extended by the right neutrino singlets. To this end we
have adapted the standard formalism of perturbative S-matrix  for
calculating the amplitudes of the processes passing at finite
distances and finite time intervals by modifying the Feynman
propagator. The developed approach is physically transparent and,
unlike  the standard one, has the advantage of not violating the
energy-momentum conservation. In its framework, the calculation of
amplitudes is carried out  in the way that is most similar to the
one used in the usual perturbative S-matrix formalism, which is
much simpler than in papers \cite{Giunti:1993se,Naumov:2010um},
where an  approach based on  the use of wave packets and  the
standard Feynman  propagators for describing the motion of virtual
neutrinos has been developed.

The application of this modified formalism to describing the
neutrino oscillation  processes with  neutrino detection by
charged-current and  neutral-current  interactions with electrons
showed that the standard results can be easily and consistently
obtained using only the mass eigenstates of these particles. This
differs the developed approach from the approaches of papers
\cite{Giunti:1993se,Grimus:1996av,Naumov:2010um,Lobanov:2015esa},
where the results of describing the neutrino oscillation processes
with  neutrino detection by charged-current interaction include
corrections to the results of the standard approach coming from
the wave packet structure of the initial particle states.

%\newpage
\bigskip
{\large \bf Acknowledgments}
\medskip \\
\noindent The authors are  grateful to E. Boos,  A. Lobanov and M.
Smolyakov for reading the manuscript and making important comments
and to L. Slad for useful discussions.   Analytical calculations
of the amplitudes have been carried out with the help of  the
COMPHEP and REDUCE packages. The work was supported by grant
NSh-7989.2016.2 of the President of Russian Federation.

\end{document}